\documentclass[conference]{IEEEtran}

 \ifCLASSINFOpdf
  \usepackage[pdftex]{graphicx}
     \DeclareGraphicsExtensions{.pdf,.jpeg,.png}
\else
     \usepackage[dvips]{graphicx}
     \DeclareGraphicsExtensions{.eps}
\fi
 
\begin{document}

\title{A Microcontroller Based Device to Reduce Phanthom Power}

\author{
\IEEEauthorblockN{Nitin Jagadish\IEEEauthorrefmark{1}, Manoj H\IEEEauthorrefmark{2} and Nishanth K Prasad \IEEEauthorrefmark{3}}
\IEEEauthorblockA{Nitte Meenakshi Institute of Technology\\
Govindapura, Gollahalli, Yelahanka,\\Bangalore - 560 064}
\IEEEauthorblockA{\IEEEauthorrefmark{1}jnitingowda@gmail.com, \IEEEauthorrefmark{2}manojh94@gmail.com, \IEEEauthorrefmark{3}nishanth.nkp@gmail.com}
\\\\
\IEEEauthorblockN{Sunil Kumar K M}
\IEEEauthorblockA{M S Ramaiah Institute of Technology, Bangalore, India 560-054\\}
\IEEEauthorblockA{sunilkumarkm8@gmail.com}
} 

\maketitle 
 
\begin{abstract} 
In this paper we concern ourselves with the problem of minimizing the standby power consumption in some of the house hold appliances. Here we propose a remote controlled device through which we could reduce the amount of standby power consumed by the electrical appliances connected to it. This device provides an option of controlling each of the appliances connected to it individually or as a whole when required. The device has got number of plug points each of which could be controlled through the remote and also has a provision of switching off all the points at once. 
\end{abstract}

\IEEEpeerreviewmaketitle

\section{Introduction} 

Standby power is the electricity consumed by end-use electrical equipment when it is switched off or not performing its main function. The most common users of standby power are televisions (TVs) and video equipment with remote controls, electrical equipment with external low-voltage power supplies, office equipment, and devices with continuous digital displays. 

The actual power draw in standby mode is small, typically 0.5–30 watts\cite{1}. However, Standby power is consumed 24 hours per day, and more and more new appliances have features that consume standby power. Although consumption by individual appliances is small, the cumulative total is significant. It was reported that standby power represents 20–60 W per home in developed countries, and is responsible for about 2\% of the total electricity consumption in OECD(Organization for Economic Cooperation and Development) countries\cite{2}.

 Recent estimates of standby use range from 3 to 10 percent of residential electricity use1, depending on the country and the specific measurement procedures used in the surveys. However, when all electronic appliances in homes and offices in a single country are aggregated, the standby power they consume represents a significant fraction of total electricity use. Reduction of standby power consumption could reduce $CO_2$ emissions as well.

A study done by Australian green house office concludes that up to 80\% of the electricity used in video recorders was in standby mode. In NewZeland, televisions consume 40\% of electricity as standby energy. Field surveys conducted in office buildings of Thailand in 1996 showed that idle losses were 53\% for personal computers, over 90\% for copiers, printers and fax machines\cite{3}.

Some of the solutions for this problem include switching off the main or unplugging the device from the mains when we are not using it. We could use a power strip with switch so that we could turn off the same, in turn disconnecting the devices connected to it. One of the disadvantages of this power strip is that all appliances connected to it get disconnected when switched off. Our proposal here is an improvised version of it allowing the user to control each of the appliances without disturbing the function of the others using a programmable remote.

\subsection*{ Operational modes of electrical appliances and power required}
\subsubsection{Operational mode}
The appliance is used as intended (i.e., performing its primary function). The appliance draws the most power in this mode. The power required when the appliance is in the normal operational mode.

\subsubsection{Active standby Mode}
The appliance is turned on, but is not actually being used (e.g., many computer monitors have this feature). The power is required when the appliance is in energy saving mode. 

\subsubsection{Passive standby mode}
The appliance (e.g., a television set) is turned off by the remote control. The power is required when the appliance is in standby mode, waiting to perform an intended function. 

\subsubsection{Off mode}
The appliance is not switched on, nor will the remote control, if any, work. The appliance is disconnected from the power source. 

\section{Our Approach}
\begin{figure}[!t]
\centering
\includegraphics[width=9cm, height=5.5cm]{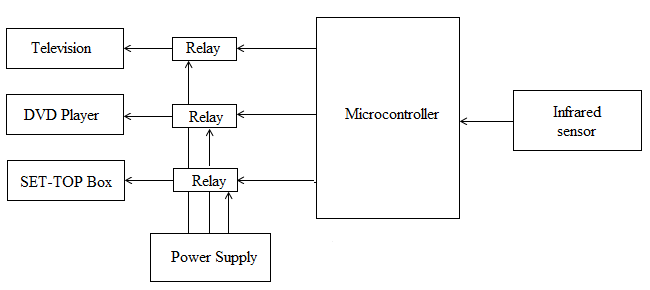}\label{1}
\caption{Working of the PowerStrip}
\end{figure}

The system proposed is programmable remote controlled and can be used to connect any of the required devices. Each of the appliances could be operated through remote separately and another option is given for the user to shut down the entire device, disconnecting all the appliances connected. \ref{1} shows the overview of the system. 

\begin{figure}[!t]
\centering
\includegraphics[width=9cm, height=5.5cm]{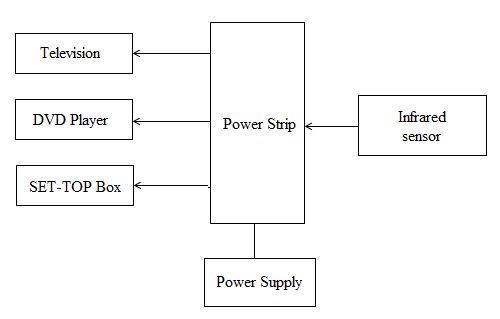}\label{2}
\caption{Overall Working of the System}
\end{figure}

The appliances are connected to the plug points in the device which is in turn connected to the main power supply. 
The electronics part of the device contains an infrared sensor, a control unit (micro controller) and relays as shown in \ref{2}.

\subsection{Operation}
The device is programmable remote controlled. It works with all kinds of remotes. One special feature about the remotes is that it can be programmed with macros, which enable us to program several commands into one button. For example, we can program a macro that turns on both our TV and satellite dish with a single press of a button. Or we can program TV and DVD player to turn on at the same time. The instruction manual that comes with the remote will contain a list of macro functions for the remote that can be programmed or downloaded from the universal remote\cite{rem} developer's Web site\cite{4}.
The plug points in the power strip are numbered and the remote is provided with separate numbered buttons for each of the plug points. Pressing the button will switch on the respective plug point in turn the appliance connected to it. To switch on the next plug point the respective numbered button is pressed. The remote is provided with another power button which switches off all the plug points and switches on when pressed again.

\subsection{Infrared detector}
The property of LEDs is that they produce a certain wavelength of light when electric current is applied but they also produce a current when they are subjected to the same wavelength light.
The infrared sensor interprets the signals as mentioned, from the remote and is given to the control unit.
The control unit is programmed to act upon different plug points based on the interpreted signals. It energizes the required relay which connects the appliance to the main power supply. 

\subsection{Relay}
A relay is used when it is required to control a high voltage circuit using a lower DC voltage. A relay is an electrically operated switch. Current flowing through the coil of the relay creates a magnetic field which attracts a lever and changes the switch contacts. The coil current can be on or off so relays have two switch positions. . Current flowing through the coil of the relay creates a magnetic field which attracts a lever and changes the switch contacts. The coil current can be on or off so relays have two switch positions.

The relay's switch connections are usually labeled COM, NC and NO: \\
\subsubsection*{COM}Common, always connect to this; it is the moving part of the switch
\subsubsection*{NC}Normally Closed, COM is connected to this when the relay coil is off. 
\subsubsection*{NO}Normally Open, COM is connected to this when the relay coil is on. \\

\begin{figure}[!t]
\centering
\includegraphics[width=9cm, height=5.5cm]{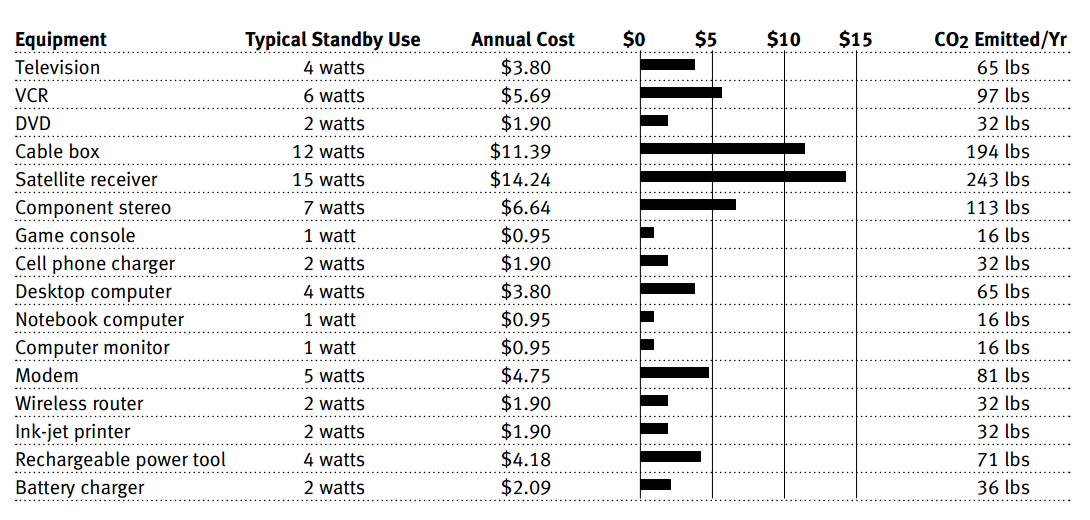}\label{3}
\caption{Statistics by \textit{responsiblenergy}}
\end{figure}

Our design has the NO pin connected to the main power supply and the COM pin connected to the appliance. When the control unit triggers the relay the NO pin and COM pin gets connected which results in switching on the appliance\cite{5}.
For the device operating on three appliances as in \ref{1} the control unit is programmed for four kinds of signals, three of them for controlling each of the three appliances and one more to shut down the device completely. Figure \ref{3} shows the statistics conducted by \textit{responsiblenergy} \cite{6}. 

\section{Conclusion} 
In this paper, we propose a method to save the phantom power. Its efficiency can be further improvised by programming the individual socket, depending on its funcionalities. The sockets' control can be made available with long range wireless medium like gsm to control it on the go. This helps in controlling the appliances as and when needed, instead of leaving it on always.

\section{Acknowledgement}
The authors would like to thank M.S.Ramaiah Institute of Technology, Prof.Vijaya Kumar.B.P( Head, Dept of ISE, M.S.Ramaiah Institute of Technology) and Nitte Meenakshi Institute of Technology for the support and motivation.

\end{document}